\documentclass[aps,prb,twocolumn,groupedaddress,floats,showpacs]{revtex4}

\usepackage{amsmath}
\usepackage{amssymb}
\usepackage[dvips]{graphics}
\usepackage[dvipdfm]{hyperref}
\usepackage{epsf}

\newcommand{\e}{\varepsilon}

\newcommand{\D}{\Delta}

\newcommand{\omegamatsubara}{\Omega}
\newcommand{\xy}{{\rm eff}}
\newcommand{\pot}{{\mu}}

\newcommand{\up}{\uparrow}
\newcommand{\dw}{\downarrow}

\newcommand{\phim}{\phi_{\rm m} }
\newcommand{\thm}{\theta_{\rm m}}

\renewcommand{\vec}[1]{\mathbf{#1}}
\renewcommand{\vr}{\vec{r}}
\newcommand{\vm}{\vec{m}}
\newcommand{\ve}{\vec{e}}
\newcommand{\vsigma}{\mbox{\boldmath $\sigma$}}
\newcommand{\vsigmascript}{\mbox{\scriptsize \boldmath $\sigma$}}
\newcommand{\vA}{\vec{A}}
\newcommand{\vnabla}{\mbox{\boldmath $\nabla$}}
\newcommand{\vk}{\vec{k}} 
\newcommand{\vR}{\vec{R}} 
\newcommand{\vkp}{\vec{k}_{\parallel}}

\newcommand{\ket}[1]{|#1\rangle}

\begin{document}

\title{Andreev reflection at half-metal--superconductor interfaces with non-uniform magnetization}

\author{Joern N. Kupferschmidt$^{1,2}$ and Piet W. Brouwer$^{2}$}
\affiliation{$^{1}$Laboratory of Atomic and Solid State Physics, Cornell
University, Ithaca, NY 14853-2501, USA \\
$^{2}$Dahlem Center for Complex Quantum Systems and Institut f\"ur theoretische Physik, Freie Universit\"at Berlin, Arnimallee 14, 14195 Berlin, Germany}
\date{\today}

\begin{abstract}
Andreev reflection at the interface between a half-metallic ferromagnet and a spin-singlet superconductor is possible only if it is accompanied by a spin flip. Here we calculate the Andreev reflection amplitudes for the case that the spin flip originates from a spatially non-uniform magnetization direction in the half metal. We calculate both the microscopic Andreev reflection amplitude for a single reflection event and an effective Andreev reflection amplitude describing the effect of multiple Andreev reflections in a ballistic thin film geometry. It is shown that the angle and energy dependence of the Andreev reflection amplitude strongly depends on the orientation of the gradient of the magnetization with respect to the interface. Establishing a connection between the scattering approach employed here and earlier work that employs the quasiclassical formalism, we connect the symmetry properties of the Andreev reflection amplitudes to the symmetry properties of the anomalous Green function in the half metal.
\end{abstract}


\maketitle

\section{Introduction}

Superconductors extend their order into adjacent normal-metals via the mechanism of Andreev reflection:\cite{kn:imry2002} An electron incident on the superconductor--normal-metal interface is phase-coherently reflected as a hole, and vice versa.\cite{kn:andreev1964} Half-metallic ferromagnets (half metals) are normal metals that support quasiparticle excitations of only one spin orientation. 
If superconducting correlations are to extend from a spin-singlet superconductor into a half-metallic ferromagnet, the azimuthal spin-rotation symmetry around the magnetization axis needs to be broken near the superconductor--half-metal interface, so that electrons can be Andreev reflected into holes in the same spin band.

In standard ferromagnets (which have majority as well as minority carriers), such a combination of spin-flip and Andreev reflection was shown theoretically\cite{kn:bergeret2001,kn:bergeret2001c,kn:kadigrobov2001,kn:bergeret2005} and experimentally \cite{kn:sosnin2006,kn:krivoruchko2007,kn:yates2007,kn:khaire2010} to lead to a superconductor proximity effect with a range comparable to that in normal metals. In the literature, magnetic domain walls,\cite{kn:bergeret2001,kn:bergeret2001c,kn:kadigrobov2001,kn:volkov2008b,kn:volkov2005,kn:fominov2007,kn:linder2009b} helical structures intrinsic to the ferromagnetic material,\cite{kn:volkov2006,kn:linder2009b,kn:halasz2009,kn:alidoust2010} artificially structured multilayers with noncollinear magnetization directions in different layers,\cite{kn:volkov2003,kn:houzet2007,kn:braude2007,kn:volkov2010,kn:trifunovic2010} and a precessing magnetization direction\cite{kn:houzet2008} were addressed as possible microscopic origins of the broken spin-rotation symmetry.
Although the possibility of spin-flip Andreev reflection (with similar origins) also exists for half-metal--superconductor interfaces,\cite{kn:eschrig2003,kn:braude2007,kn:asano2007a,kn:asano2007b,kn:eschrig2008,kn:galaktionov2008,kn:beri2009} and a sizeable Josephson current has been observed in junctions involving a half-metallic ferromagnet,\cite{kn:keizer2006,kn:anwar2010} the Andreev reflection mechanism at half-metal--superconductor interfaces was found to be more delicate than in the case of standard ferromagnets.\cite{kn:beri2009} 

The underlying reason for the differences between a ferromagnet--superconductor (FS) interface and a half-metal--superconductor (HS) interface is that, if no orbital symmetries are broken, the latter admits a description in terms of an effective $2 \times 2$ scattering matrix, whereas the former requires at least a $4 \times 4$ scattering matrix, to account for the spin degree of freedom. Mathematical constraints on $2 \times 2$ matrices following from current conservation and particle-hole symmetry then force the Andreev reflection amplitude $r_{\rm he}(\varepsilon)$ of an HS junction to be generically zero at the Fermi level $\varepsilon=0$, whereas there is no such strong restriction on the Andreev reflection amplitudes for an FS junction. As shown in Ref.\ \onlinecite{kn:beri2009}, the fact that $r_{\rm he}(0)=0$ for an HS junction has little consequences for the strength of the proximity effect at distances below the superconducting coherence length $\xi_{\rm S}$, but leads to a suppression of the proximity effect at larger distances in comparison to a normal-metal--superconductor (NS) junction.\cite{kn:beri2009} For example, the zero-temperature critical current in a ballistic 
SHS
junction of length $L \gg \xi_{\rm S}$ is a factor $\sim (L/\xi_{\rm S})^2$ smaller than in an otherwise comparable
SFS  
junction.\cite{kn:beri2009} (The experiments of Refs.\ \onlinecite{kn:keizer2006,kn:anwar2010}, which observe a sizeable Josephson effect in SHS junctions, have $L \sim \xi_{\rm S}$, so that they need not be affected by this effect.)

If orbital symmetries are broken, the mathematical constraints leading to the condition $r_{\rm he}(\varepsilon)=0$ at $\varepsilon=0$ are no longer operative. Examples of broken orbital symmetries that allow for a finite Andreev reflection amplitude $r_{\rm he}(\varepsilon)$ at $\varepsilon=0$ are a magnetization gradient parallel to the interface, spin-orbit coupling in the superconductor --- which both break inversion symmetry with respect to the interface normal ---, or impurity scattering. Andreev reflection in the presence of a magnetization gradient parallel to the interface, which is relevant for a domain wall in the half metal, was previously considered by us in Ref.\ \onlinecite{kn:kupferschmidt2009} in the limiting case of a half metal with infinite wavefunction decay rate $\kappa_{\downarrow}$ of the minority carriers. This situation will be analyzed in more detail here. We lift simplifying assumptions of Ref.\ \onlinecite{kn:kupferschmidt2009}, and compare the generalized result to the case when the magnetization varies perpendicular to the interface, where no orbital symmetry is broken. The other two scenarios for breaking orbital symmetries will be analyzed elsewhere.\cite{kn:duckheim2010,kn:wilken2010}

\begin{figure}
\bigskip
\epsfxsize=0.9\hsize
\epsffile{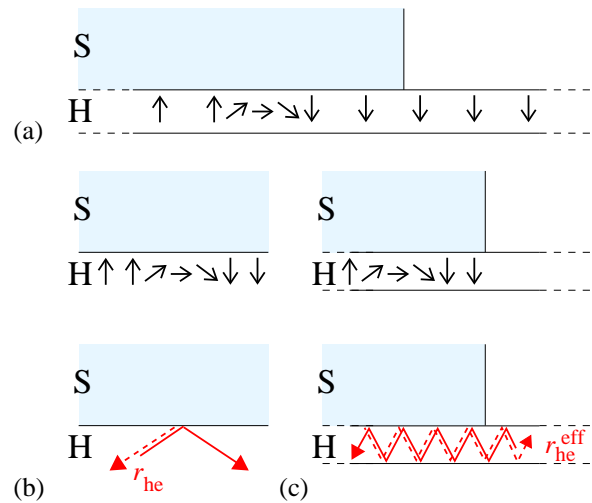}
\caption{\label{fig:1} (Color online) (a) Lateral contact between a thin half-metallic film (H) and a superconductor (S). Andreev reflection at the HS interface is possible at those interface positions where the magnetization direction (indicated by thin black arrows) is not uniform, such as a domain wall. The situation shown in the figure is generic, since domain walls at the HS interface are bound to occur if the HS contact is larger than the domain size in the half-metallic film. In this geometry, there are two possible ways to define the Andreev reflection amplitude: The amplitude $r_{\rm he}$ for a single Andreev reflection (thick red arrows), appropriate for the interface between a semi-infinite half metal and a superconductor (b), and the effective amplitude $r_{\rm he}^{\rm eff}$ representing the combined effect of multiple Andreev reflections in the same region of non-uniform magnetization for quasiparticles moving in the half metallic film (c). In parts (b) and (c), the black arrows in the top panel show the spatial variation of the magnetization direction, whereas red arrows in the bottom panel show an example of a trajectory for an incoming electron (solid) and an Andreev reflected hole (dashed).}
\end{figure}

The scenario of a magnetization gradient parallel to the HS interface is particularly relevant for a lateral superconductor--half-metal contact, for which the superconducting contact is deposited on top of a half-metallic film. The experiments of Refs.\ \onlinecite{kn:keizer2006,kn:anwar2010} were performed in such a lateral geometry. A lateral contact, with a domain wall in the half metal under the contact area, is shown schematically in Fig.\ \ref{fig:1}.
For a phase-coherent lateral contact, it is convenient to define an ``effective'' Andreev reflection amplitude $r_{\rm he}^{\xy}$ that represents the effect of multiple Andreev reflections at the HS interface for a quasiparticle moving in the thin half-metallic film, incident on a domain wall (or any other region with non-uniform magnetization) from a region with uniform magnetization, see Fig.\ \ref{fig:1}c. (The standard definition of the Andreev reflection amplitude $r_{\rm he}$ is for single incidence of a quasiparticle onto the superconductor interface. This is the situation shown in Fig.\ \ref{fig:1}b.) Because one is interested in currents flowing parallel to the interface in a lateral contact (see Fig.\ \ref{fig:1}c), it is the effective amplitude $r_{\rm he}^{\xy}$, not $r_{\rm he}$, which occurs in a calculation of, {\em e.g.}, the subgap conductance or the Josephson current.\cite{kn:kupferschmidt2009} 

In this article, we calculate both the microscopic Andreev reflection amplitudes $r_{\rm he}$ and the effective amplitude $r_{\rm he}^{\xy}$. We consider an HS interface with a low normal-state transmission probability, caused by the presence of a tunnel barrier or by the mismatch of Fermi velocities on both sides of the junction. We do not consider the effect of impurity scattering. For the calculation of $r_{\rm he}$ this is not a serious shortcoming as Andreev reflection is local, and only impurities in the immediate vicinity of the interface would play a role (up to distances of the order of a Fermi wavelength). For the effective Andreev reflection amplitude $r_{\rm he}^{\xy}$ in the thin-film geometry, the omission of impurity scattering limits the applicability of our results to situations in which the mean free path is larger than the domain wall size $l_{\rm d}$.
In addition to the scenario of a magnetization gradient parallel to the HS interface, we also consider the case that the magnetization gradient is perpendicular to the HS surface as is relevant, {\em e.g.}, if the half metal has different boundary and bulk anisotropies.

The calculation of the microscopic, single-reflection amplitude $r_{\rm he}$ will be presented first. Hereto, we first review the restrictions of symmetry on Andreev reflection at a HS interface in Sec.\ \ref{sec:symcons}. We then introduce the Hamiltonian and the wavefunctions for the case of a uniform magnetization in Sec.\ \ref{sec:model}. The calculation of the Andreev reflection amplitudes in the presence of a non-uniform magnetization then takes place in Sec.\ \ref{sec:overlap}. The calculation of the effective Andreev reflection amplitude $r_{\rm he}^{\xy}$ for a superconductor placed on top of a thin half-metallic film is then presented in Sec.\ \ref{sec:wg}. Finally, in Sec.\ \ref{sec:green}, we discuss the relation of the scattering approach used in this article and the Green function approach used in most of the literature. In particular, we consider the anomalous Green function and its frequency dependence. We conclude in Sec.\ \ref{sec:conc}. The appendix contains certain details of the calculations not presented in the main text.


\section{Constraints imposed by unitarity and particle-hole degeneracy} 
\label{sec:symcons}

\begin{figure}
\bigskip
\epsfxsize=0.8\hsize
\epsffile{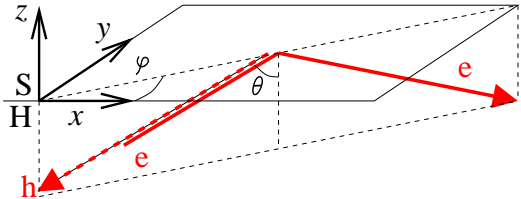}
\caption{\label{fig:2} (Color online) Illustration of the scattering setup. Coordinates are chosen, such that the half-metal--superconductor interface is the plane $z=0$. Electron-like quasiparticles incident on the HS interface are either reflected as an electron (normal reflection), or as a hole (Andreev reflection). At a translationally invariant interface, the projection $\vkp$ of the wavevector on the interface is conserved upon reflection.}
\end{figure}

Although the considerations of this section are completely general, for definiteness we choose coordinates such that the half-metal--superconductor interface is the plane $z=0$. In this section and in the following two sections, the half metal occupies the half space $z < 0$, and the superconductor, which is taken to be of $s$-wave, spin-singlet type, occupies the half space $z > 0$. This setup is shown schematically in Fig.\ \ref{fig:2}.

Quasiparticles in the half metal are labeled by their wavevector $\vkp = k_x \ve_x + k_y \ve_y$ parallel to the HS interface ($\ve_x$ and $\ve_y$ are unit vectors in the $x$ and $y$ directions, respectively) and by their excitation energy $\varepsilon$. 
We assume periodic boundary conditions in the $x$ and $y$ directions, so that the wavevectors $\vkp$ are discrete. 

At excitation energies $\varepsilon$ below the superconducting gap $\Delta$, quasiparticles incident on the half-metal--superconductor interface from the half-metallic side will be reflected back into the half metal. This reflection can be either normal reflection or Andreev reflection, for which electron-like quasiparticles are reflected as holes and vice versa. The reflection process is described by a scattering matrix ${\cal S}(\vkp',\vkp;\varepsilon)$, which takes the form
\begin{equation}
  {\cal S}(\vkp',\vkp;\varepsilon) =
  \left( \begin{array}{cc} 
  r_{\rm ee}(\vkp',\vkp;\varepsilon)  &  r_{\rm eh}(\vkp',\vkp;\varepsilon)  \\
  r_{\rm he}(\vkp',\vkp;\varepsilon) &  r_{\rm hh}(\vkp',\vkp;\varepsilon)
  \end{array} \right), 
\end{equation}
where 
the subscripts e and h refer to electron-like and hole-like states. 
The scattering amplitudes $r_{\rm eh}$ and $r_{\rm he}$ describe Andreev reflection processes. 

The scattering matrix ${\cal S}(\vkp',\vkp;\varepsilon)$ satisfies two constraints: Unitarity and particle-hole symmetry. The latter condition reads
\begin{equation}
  {\cal S}(\vkp',\vkp;\varepsilon) =
  \left( \begin{array}{cc} 0 & 1 \\ 1 & 0 \end{array} \right)
  {\cal S}(-\vkp',-\vkp;-\varepsilon)^*
  \left( \begin{array}{cc} 0 & 1 \\ 1 & 0 \end{array} \right).
  \label{eq:phsymmetry}
\end{equation}
The combination of unitarity and particle-hole degeneracy severely restricts the form of the scattering matrix if there is translation invariance along the interface, which implies 
\begin{equation}
  {\cal S}(\vkp',\vkp;\varepsilon) = {\cal S}(\vkp;\varepsilon) \delta_{\vkp',\vkp}. \label{eq:translation}
\end{equation}
Combination of Eqs.\ (\ref{eq:phsymmetry}) and (\ref{eq:translation}) gives
\begin{eqnarray}
  r_{\rm ee}(\vkp,\varepsilon) &=& r_{\rm hh}(-\vkp,-\varepsilon)^*, \nonumber \\
  r_{\rm eh}(\vkp,\varepsilon) &=& r_{\rm he}(-\vkp,-\varepsilon)^*.
\end{eqnarray}
If the scattering problem is also invariant for a $\pi$ rotation around the interface normal, so that
\begin{equation} 
  {\cal S}(\vkp',\vkp;\varepsilon) = {\cal S}(-\vkp',-\vkp;\varepsilon),
  \label{eq:inversion}
\end{equation}
one finds, upon combining all symmetry properties, 
\begin{equation}
  r_{\rm ee}(\vkp;0) r_{\rm eh}(\vkp;0) = 0
\end{equation}
for the scattering matrix at the Fermi level $\varepsilon = 0$.
Since $r_{\rm ee} \neq 0$, except for a special choice of parameters, this implies that generically one must have\cite{kn:beri2009}
\begin{equation}
  r_{\rm eh}(\vkp;0) = 0. \label{eq:zero}
\end{equation}

These very general considerations pose a severe restriction on the magnitude and the spatial extension of the proximity effect in half metals that is absent in ferromagnet--superconductor junctions with otherwise comparable characteristics. A nonzero Andreev reflection amplitude for a half-metal--superconductor junction can be obtained only by fine-tuning device parameters such that the normal reflection amplitude becomes zero, or by invoking processes that break the symmetries leading to Eq.\ (\ref{eq:zero}). The former scenario was discussed in Ref.\ \onlinecite{kn:beri2009} and will not be addressed here. Examples of symmetry-breaking processes that result in a nonzero Andreev reflection amplitude are: lifting of particle-hole degeneracy by a finite excitation energy $\varepsilon$,\cite{kn:beri2009,kn:eschrig2009} breaking of the rotation symmetry around the interface normal,\cite{kn:kupferschmidt2009} breaking of the translation symmetry along the interface, or the breaking of phase coherence.\cite{kn:beri2009b} A domain wall for which the magnetization direction varies in a direction parallel to the interface is an example of a perturbation that breaks the rotation symmetry.\cite{kn:kupferschmidt2009} However, a thin interface layer of different magnetic orientation than the interior of the half metal (which is a model of a ``spin-active interface''\cite{kn:eschrig2003,kn:asano2007a,kn:asano2007b,kn:eschrig2008,kn:kalenkov2009}) does not lift the constraints leading to Eq.\ (\ref{eq:zero}).\cite{kn:beri2009,kn:eschrig2009} The role of variations in the magnetization direction will be considered in more detail in Sec.\ \ref{sec:overlap} below.

A finite excitation energy $\varepsilon$ lifts the particle-hole degeneracy, and the Andreev reflection amplitude $r_{\rm eh}$ becomes nonzero. If no other symmetries are broken, the order of magnitude of the Andreev reflection amplitudes at finite $\varepsilon$ can be estimated as
\begin{equation}
  |r_{\rm eh}(\varepsilon)| \sim \frac{|\varepsilon|}{\min(\Delta/\tau,E_{\xi})} |r_{\rm eh,\, FS}|,
\end{equation}
where $E_{\xi}$ is the Thouless energy of the interface layer where the singlet-triplet conversion takes place, $\tau$ is the transparency of the superconductor interface, $\Delta$ the superconducting gap, and $r_{\rm eh,\, FS}$ the Andreev reflection amplitude of a ferromagnet--superconductor amplitude of otherwise comparable characteristics. The first energy scale in the denominator comes about because electrons and holes scattering off a normal-metal--superconductor interface of transparency $\tau$ at finite excitation energy $\varepsilon$ experience an additional phase difference $\sim \pm \varepsilon \tau/\Delta$, which lifts the electron-hole degeneracy.\cite{kn:blonder1982} The second energy scale in the denominator appears from phase differences acquired in the interface layer. The typical thickness of this interface layer is of the order of minority decay length $\xi$, which implies that $E_{\xi}$ is of the order of the Fermi energy. For tunneling interfaces one always has $|r_{\rm eh}(\varepsilon)| \ll |r_{\rm eh,\, FS}|$ and we conclude that the breaking of electron hole symmetry by finite excitation energies is not an efficient route towards sizeable Andreev reflection in that case. The suppression of Andreev reflection in half-metal--superconductor junctions (as compared to ferromagnet--superconductor junctions) is absent only for transparent interfaces and excitation energies of order $\Delta$.

The $\varepsilon$-dependence of $r_{\rm eh}$ not only determines the conductance through the half-metal--superconductor interface at finite bias,\cite{kn:beri2009,kn:eschrig2009} it also sets the scale for the Josephson effect in a superconductor-half-metal--superconductor junction.\cite{kn:beenakker1991d,kn:brouwer1997e} If the Thouless energy $E_{L}$ of a Josephson junction of length $L$ is large in comparison to $\Delta$ (``short junction limit''), the Josephson current $I$ is carried by quasiparticle states with energies up to $\Delta$. In this limit, the symmetry considerations that suppressed Andreev reflection at $\varepsilon=0$ do not affect the order of magnitude of $I$, and one concludes that otherwise comparable superconductor-half-metal--superconductor and superconductor--ferromagnet--superconductor junctions have comparable Josephson currents.\cite{kn:beri2009} If, however, $E_{L} \ll \Delta$ (``long junction limit''), only quasiparticle states with energy below $E_{L}$ contribute to $I$, so that $I$ is significantly suppressed compared to the Josephson current in comparable superconductor--ferromagnet--superconductor junctions.

In the remainder of this article, we present explicit model calculations of the Andreev reflection amplitudes for the case that singlet-triplet conversion is mediated by a spatially non-uniform magnetization in the half metal, as it appears, e.g., in a domain wall.

\section{Hamiltonian and Scattering states} 
\label{sec:model}

\subsection{Bogoliubov-de Gennes Hamiltonian}

Quasiparticle excitations near the HS interface are described by the Bogoliubov-de Gennes equation\cite{kn:beenakker1995}
\begin{equation} 
\label{eq:BdG}
  {\cal H}
 \Psi(\vr)
  = \e \Psi(\vr),
  \ \ 
  {\cal H} = \left( \begin{array}{cc} 
  \hat H   &   i \Delta e^{i \phi} \sigma_2 \\
- i \Delta e^{-i \phi} \sigma_2 & - \hat H^*  
\end{array} \right),
\end{equation}
where the four-component spinor
\begin{equation} 
 \Psi(\vr)  = ( u_{\up }(\vr) , u_{\dw}(\vr) , v_{\up}(\vr) , v_{\dw}(\vr) )^{\rm T} 
\end{equation}
consists of wavefunctions $u_{\sigma}(\vr) $ for the electron and $v_{\sigma}(\vr)$ for the hole degrees of freedom. The superconducting order parameter
$\Delta (\vr) e^{i \phi}$ is nonzero only in the superconductor.
We will take $\Delta (\vr) = \Delta \Theta(z)$, where $\Theta(z) = 1$ if $z > 0$ and $0$ otherwise. This step function model is appropriate for tunneling interfaces of $s$-wave superconductors.\cite{kn:likharev1979}

For the single-particle Hamiltonian, we take the simplest model that contains the essential features of the half-metal--superconductor interface,
\begin{eqnarray}
  \hat H &=&
  -\hbar^2 \vnabla \frac{1}{2 m(z)} \vnabla -
  \sum_{\sigma} \pot_{\sigma}(z) \hat P_{\sigma}(\vr) + \hbar w \delta(z), \nonumber \\
\end{eqnarray}
where 
\begin{equation}
  m(z) = \left\{ \begin{array}{ll}
  m_{\rm H} & \mbox{if $z < 0$}, \\
  m_{\rm S} & \mbox{if $z > 0$}, \end{array} \right.
\end{equation}
with $m_{\rm H}$ and $m_{\rm S}$ being the effective masses for the half metal and the superconductor, respectively, 
\begin{equation}
  \pot_{\sigma}(z) = \left\{ \begin{array}{ll}
  \pot_{{\rm H}\sigma} & \mbox{if $z < 0$}, \\
  \pot_{{\rm S}} & \mbox{if $z > 0$}, \end{array} \right.
\end{equation}
with $\sigma=\uparrow,\downarrow$ and the potentials $\pot_{{\rm H}\uparrow}$, $\pot_{{\rm H}\downarrow}$, and $\pot_{\rm S}$ representing the combined effect of the chemical potential and band offsets for the majority and minority electrons in the half metal and for the superconductor, respectively, and where $w$ sets the strength of a delta-function potential barrier at the interface. The operators 
\begin{equation}
  \hat P_{\uparrow} = \frac{1}{2} + \frac{1}{2} \vm(\vr) \cdot \hat{\vsigma}, \ \
  \hat P_{\downarrow} = \frac{1}{2} - \frac{1}{2} \vm(\vr) \cdot \hat{\vsigma}
\end{equation}
project onto the majority and minority components, respectively, where 
$\vm(\vr)$ is a unit vector pointing along the magnetization direction in the half metal. 

The potentials $\pot_{{\rm H}\uparrow}$, $\pot_{{\rm H}\downarrow}$, and $\pot_{\rm S}$ are chosen such that $\pot_{{\rm H}\uparrow}$, $\pot_{{\rm S}} > 0$, and $\pot_{{\rm H}\downarrow} < 0$. As a result, majority states in the half metal and in the normal state of the superconductor are propagating states, with Fermi wavenumbers 
\begin{equation}
  k_{\uparrow} = \frac{1}{\hbar}\sqrt{2 m_{\rm H} \pot_{{\rm H}\uparrow}},\ \
  k_{\rm S} = \frac{1}{\hbar}\sqrt{2 m_{\rm S} \pot_{\rm S}},
\end{equation}
respectively. The corresponding Fermi velocities are $v_{\uparrow} = \hbar k_{\uparrow}/m_{\rm H}$ and $v_{\rm S} = \hbar k_{\rm S}/m_{\rm S}$, respectively.
Minority states in the half metal are evanescent with wavefunction decay rate
\begin{equation}
  \kappa_{\downarrow} = \frac{1}{\hbar} \sqrt{2 m_{\rm H} |\pot_{{\rm H}\downarrow}|}.
\end{equation}
(The wavefunction decay rate $\kappa_{\downarrow}$ is the inverse of the wavefunction decay length $\xi$ used in the previous section.)
The strength $w$ of the $\delta$-function potential is chosen such that the transmission probability of the interface is much smaller than unity. It is in this limit only, that the step-function model for the superconducting order parameter $\Delta$ used in Eq.\ (\ref{eq:BdG}) is valid.\cite{kn:likharev1979} We use the Andreev approximation $\Delta \ll \mu_{\rm S}$ throughout our calculation.


\subsection{Scattering states for $\Delta = 0$}

In order to introduce the relevant notation, we first consider solutions of the Bogoliubov-de Gennes equation (\ref{eq:BdG}) in the normal state ({\em i.e.}, with $\Delta =0$), for a spatially uniform magnetization direction $\vm = \ve_3$, and at $\varepsilon=0$. In this case, solutions of the Bogoliubov-de Gennes equation (\ref{eq:BdG}) can be written as a product
\begin{equation}
  \Psi(\vec{\vr}) = e^{i \vec{k}_{\parallel} \cdot \vr } 
  \Psi_{\vkp}(z), 
\end{equation}
where $\vkp = k_x \ve_x + k_y \ve_y$.
For $z < 0$, the spinor wavefunction $\Psi_{\vkp}$ has the general form
\begin{eqnarray}
  \Psi_{\vkp}(z) &=&
  \frac{1}{\sqrt{v_{\uparrow z}}}
  \left( \begin{array}{c} 
  c_{{\rm e}\uparrow} e^{i k_{\uparrow z} z} +
    c_{{\rm e}\uparrow}' e^{-i k_{\uparrow z} z} \\
  0 \\
  c_{{\rm h}\uparrow} e^{-i k_{\uparrow z} z} +
  c_{{\rm h}\uparrow}' e^{i k_{\uparrow z} z} \\ 0
  \end{array} \right)
  \nonumber \\ && \mbox{}
  +
  \frac{1}{\sqrt{v_{\downarrow z}}}
  \left( \begin{array}{c} 
  0 \\ c_{{\rm e}\downarrow} e^{\kappa_{\downarrow z} z} \\
  0 \\ c_{{\rm h}\downarrow} e^{\kappa_{\downarrow z} z}
  \end{array} \right),
\end{eqnarray}  
where
\begin{equation}
  k_{\uparrow z} = \sqrt{k_\uparrow^2 - |\vkp|^2},\ \
  \kappa_{\downarrow z} = \sqrt{\kappa_\downarrow^2 + |\vkp|^2},
\end{equation}
and
\begin{equation}
  v_{\uparrow z} = \hbar k_{\uparrow z}/m_{\rm H},\ \
  v_{\downarrow z} = \hbar \kappa_{\downarrow z}/m_{\rm H} .
\end{equation}
In the superconductor, for $z > 0$, the general form of the spinor wavefunction is
\begin{eqnarray}
  \Psi_{\vkp}(z) &=&
  \frac{1}{\sqrt{v_{{\rm S}z}}}
  \left( \begin{array}{c} 
  d_{{\rm e}\uparrow} e^{-i k_{{\rm S}z} z} +
    d_{{\rm e}\uparrow}' e^{i k_{{\rm S}z} z} \\
  d_{{\rm e}\downarrow} e^{-i k_{{\rm S}z} z} +
    d_{{\rm e}\downarrow}' e^{i k_{{\rm S}z} z} \\
  d_{{\rm h}\uparrow} e^{i k_{{\rm S}z} z} +
    d_{{\rm h}\uparrow}' e^{-i k_{{\rm S}z} z} \\
  d_{{\rm h}\downarrow} e^{i k_{{\rm S}z} z} +
    d_{{\rm h}\downarrow}' e^{-i k_{{\rm S}z} z}
  \end{array} \right),
\end{eqnarray}
where
\begin{eqnarray}
  k_{{\rm S}z} = \sqrt{k_{\rm S}^2 - |\vkp|^2},\ \ v_{{\rm S}z} = \hbar k_{{\rm S}z}/m_{\rm S}.
\end{eqnarray}
The amplitudes appearing in the above equations are related as
\begin{eqnarray}
  \left( \begin{array}{c}
  c_{{\rm e}\uparrow}' \\
  d_{{\rm e}\uparrow}' \\
  c_{{\rm h}\uparrow}' \\
  d_{{\rm h}\uparrow}' \end{array} \right) &=&
  \left( \begin{array}{cccc}
  r & t & 0 & 0 \\
  t & r' & 0 & 0 \\
  0 & 0 & r^* & t^* \\
  0 & 0 & t^* & r'^* \end{array} \right)
  \left( \begin{array}{c}
  c_{{\rm e}\uparrow} \\
  d_{{\rm e}\uparrow} \\
  c_{{\rm h}\uparrow} \\
  d_{{\rm h}\uparrow} \end{array} \right), \nonumber \\
  \left( \begin{array}{c}
  c_{{\rm e}\downarrow} \\  
  d_{{\rm e}\downarrow}' \\  
  c_{{\rm h}\downarrow} \\  
  d_{{\rm h}\downarrow}' \end{array} \right) &=&
  \left( \begin{array}{cc}
  t_\downarrow & 0 \\ r_\downarrow' & 0 \\
  0 & t_\downarrow^* \\ 0 & r_\downarrow'^* \end{array} \right)
  \left( \begin{array}{c}
  d_{{\rm e}\downarrow} \\ d_{{\rm h}\downarrow} \end{array} \right),
\end{eqnarray}
with 
\begin{eqnarray}
  t &=& \frac{2 \sqrt{v_{\uparrow z} v_{{\rm S}z}}}{2 i w + v_{\uparrow z} + v_{{\rm S}z}}, \nonumber \\
  r &=& -1 + t \sqrt{v_{\uparrow z}/ v_{{\rm S}z}},  \nonumber \\
  r' &=& -1 + t \sqrt{v_{{\rm S}z}/v_{\uparrow z}}, \nonumber  \\
  r_\downarrow' &=&
  -1 + t_\downarrow \sqrt{v_{{\rm S}z}/v_{\downarrow z}} ,  \nonumber \\ 
  t_\downarrow &=&
  \frac{2 \sqrt{v_{\downarrow z} v_{{\rm S}z}}}{2 i w + i v_{\downarrow z} + v_{{\rm S}z}}.
\end{eqnarray}
The amplitudes $r$, $r'$, and $t$ are majority electron reflection and transmission amplitudes of the half-metal--superconductor interface (with the superconductor in the normal state); the amplitude $r_\downarrow'$ is the minority electron reflection amplitude. The coefficient $t_\downarrow$ parameterizes the evanescent wave amplitude for minority electrons in the half metal. (There is no transmission amplitude for minority electrons.) In terms of these amplitudes, the assumption of a tunneling interface translates to $|t|$, $|t_{\downarrow}| \ll 1$.

\subsection{Scattering states for uniform magnetization}

We now use the notation established in the previous subsection to construct retarded and advanced scattering states for the half-metal--superconductor interface at finite excitation energy $\varepsilon$. The scattering states will be used for the perturbation-theory calculation of the Andreev reflection amplitudes for a non-uniform magnetization in the next section.

As before, we consider a spatially uniform magnetization direction $\vm = \ve_3$. 
For each wavevector $\vkp$, there is an electron-like and a hole-like scattering state, which we label $\ket{\vkp,{\rm e}}^{{\rm R}}$ and $\ket{\vkp,{\rm h}}^{{\rm R}}$ for the retarded states and $\ket{\vkp,{\rm e}}^{{\rm A}}$ and $\ket{\vkp,{\rm h}}^{{\rm A}}$ for the advanced states. A retarded scattering state is called ``electron-like'' or ``hole-like'' if the incoming part is electron-like or hole-like, respectively, whereas one considers the outgoing part of the wavefunction for an advanced scattering state. In general, the outgoing part of a retarded scattering state will be of mixed electron/hole type, as well as the incoming part of an advanced scattering state. For the states constructed below, however, there is no mixing between electron-like and hole-like parts in the propagating components, since there is no Andreev reflection at an HS interface if the magnetization is uniform. Only the evanescent parts of the scattering states will be of mixed electron/hole type.

The scattering states are constructed from the general form of the spinor wavefunction $\Psi_{\vkp}(z)$ at finite excitation energy $\varepsilon$ and with nonzero superconducting order parameter $\Delta$, which reads
\begin{eqnarray}
  \Psi_{\vkp}(z) &=&
  \frac{1}{\sqrt{v_{\uparrow z}}}
  \left( \begin{array}{c} 
  c_{{\rm e}\uparrow} e^{i k_{\uparrow z}(\varepsilon) z} +
    c_{{\rm e}\uparrow}' e^{-i k_{\uparrow z} (\varepsilon) z} \\
  0 \\
  c_{{\rm h}\uparrow} e^{-i k_{\uparrow z}(-\varepsilon) z} +
  c_{{\rm h}\uparrow}' e^{i k_{\uparrow z}(-\varepsilon) z} \\ 0
  \end{array} \right)
  \nonumber \\ && \mbox{}
  +
  \frac{1}{\sqrt{v_{\downarrow z}}}
  \left( \begin{array}{c} 
  0 \\ c_{{\rm e}\downarrow} e^{\kappa_{\downarrow z}(\varepsilon) z} \\
  0 \\ c_{{\rm h}\downarrow} e^{\kappa_{\downarrow z}(-\varepsilon) z}
  \end{array} \right),
  \label{eq:psi1}
\end{eqnarray}    
for $z<0$, and
\begin{eqnarray}
  \Psi_{\vkp}(z) &=& \frac{e^{i k_{{\rm S}z} z -\kappa_{{\rm S}z}(\varepsilon)z}}{\sqrt{v_{{\rm S}z}}}
  \left( \begin{array}{c}
  d_\uparrow' \\
  d_\downarrow' \\
  - d_\downarrow' e^{- i \eta(\varepsilon) - i \phi}\\
  d_\uparrow' e^{- i \eta(\varepsilon) - i \phi}
  \end{array} \right)
  \nonumber \\ && \mbox{} +
  \frac{e^{-i k_{{\rm S}z} z - \kappa_{{\rm S}z}(\varepsilon)z}}{\sqrt{v_{{\rm S}z}}}
  \left( \begin{array}{c}
  d_\uparrow  \\ 
  d_\downarrow  \\
  - d_\downarrow e^{+ i \eta(\varepsilon) - i \phi}  \\  
  d_\uparrow  e^{+ i \eta(\varepsilon) - i \phi} 
  \end{array} \right)  ~~~~
  \label{eq:psi2}
\end{eqnarray}
for $z > 0$. Here we defined
\begin{eqnarray}
  k_{\uparrow z}(\varepsilon) &=& k_{\uparrow z} + \varepsilon/\hbar v_{\uparrow z}, \label{eq:kup} \\
  \kappa_{\downarrow z}(\varepsilon) &=& \kappa_{\downarrow z} - \varepsilon/\hbar v_{\downarrow z}, \label{eq:kdown} \\
  \eta(\varepsilon) &=&  \arccos(\e/\D),\\
  \kappa_{{\rm S}z}(\varepsilon) &=& (\sqrt{\D^2-\e^2})/ \hbar v_{{\rm S}z}.
  \label{eq:kS}
\end{eqnarray}
(In Eqs.\ (\ref{eq:kup}), (\ref{eq:kdown}), and (\ref{eq:kS}) we kept the leading terms in a small-$\varepsilon$ expansion only.)
Solution of the Bogoliubov-de Gennes equation (\ref{eq:BdG}) in the Andreev approximation then gives the following relations between the coefficients
\begin{eqnarray}
  c_{{\rm e}\uparrow}' &=& \left( r + \frac{t^2}{e^{2 i \eta(\varepsilon)} r_\downarrow' - r'} \right) c_{{\rm e}\uparrow}, \nonumber \\
  c_{{\rm h}\downarrow} &=& 
  \frac{t t_\downarrow e^{ i \eta(\varepsilon) - i \phi} }{e^{2 i \eta(\varepsilon)} r_\downarrow' - r'} c_{{\rm e}\uparrow},  \nonumber  \\
  d_\uparrow &=& \frac{t }{e^{2 i \eta(\varepsilon)} r_\downarrow' - r'} c_{{\rm e}\uparrow},  \nonumber  \\
  d_\uparrow' &=& \frac{t r_\downarrow' e^{2 i \eta(\varepsilon) }}{e^{2 i \eta(\varepsilon)} r_\downarrow' - r'} c_{{\rm e}\uparrow}, \nonumber \\
  c_{{\rm h}\uparrow}' &=& \left( r^* + \frac{t^{*2}}{e^{2i \eta(\varepsilon)} r_\downarrow'^* - r'^*} \right) c_{{\rm h}\uparrow}, \nonumber  \\
  c_{{\rm e}\downarrow} &=& - \frac{t^* t_\downarrow^* e^{ i \eta(\varepsilon) + i \phi}}{e^{2 i \eta(\varepsilon)} r_\downarrow'^* - r'^*} c_{{\rm h}\uparrow},  \nonumber \\
  d_\downarrow &=&  - \frac{t^* r_\downarrow'^* e^{  i \eta(\varepsilon)  + \phi }}{e^{2 i \eta(\varepsilon)} r_\downarrow'^* - r'^*} c_{{\rm h}\uparrow}, \nonumber \\
  d_\downarrow' &=& - \frac{t^* e^{ i \eta(\varepsilon) + i \phi }}{e^{ 2i \eta(\varepsilon)} r_\downarrow'^* - r'^*} c_{{\rm h}\uparrow}.
  \label{eq:ccddeq} 
\end{eqnarray}

With the help of the above wavefunctions, we construct the retarded scattering states $\ket{\vkp,{\rm e}}^{{\rm R}}$ and $\ket{\vkp,{\rm h}}^{{\rm R}}$ as the state with wavefunction
\begin{eqnarray} 
  \langle \vr \ket{\vkp,{\rm e}}^{{\rm R}} &=&
  \frac{1}{\sqrt{W_x W_y}}
  \Psi_{\vkp,{\rm e}}(z) e^{i \vkp \cdot \vr}, \nonumber \\
  \langle \vr \ket{\vkp,{\rm h}}^{{\rm R}} &=&
  \frac{1}{\sqrt{W_x W_y}}
  \Psi_{\vkp,{\rm h}}(z) e^{i \vkp \cdot \vr},
  \label{eq:psik}
\end{eqnarray}
where $W_x$ and $W_y$ are the dimensions of the HS interface in the $x$ and $y$ directions, the spinor wavefunction $\Psi_{\vkp,{\rm e}}(z)$ is given by Eqs.\ (\ref{eq:psi1}) and (\ref{eq:psi2}) with $c_{{\rm e}\uparrow} = 1$, $c_{{\rm h}\downarrow} = 0$, all other coefficients being determined by Eqs.\ (\ref{eq:ccddeq}), and the spinor wavefunction $\Psi_{\vkp,{\rm h}}(z)$ is given by Eqs.\ (\ref{eq:psi1}) and (\ref{eq:psi2}) with $c_{{\rm e}\uparrow} = 0$, $c_{{\rm h}\downarrow} = 1$. Similarly, the advanced scattering states $\ket{\vkp,{\rm e}}^{{\rm A}}$ and $\ket{\vkp,{\rm h}}^{{\rm A}}$ are then defined as the states for which $c_{{\rm e}\uparrow}'= 1$, $c_{{\rm h}\downarrow}'= 0$ and $c_{{\rm e}\uparrow}'= 0$, $c_{{\rm h}\downarrow}'= 1$, respectively, again with all other coefficients determined by Eqs.\ (\ref{eq:ccddeq}). The factors $1/\sqrt{W_x W_y}$ in Eq.\ (\ref{eq:psik}) ensure that retarded and advanced scattering states are normalized to unit incoming or outgoing flux in the $z$ direction, respectively. The lengths $W_x$ and $W_y$ do not appear in the final expressions for the Andreev reflection amplitudes.
These scattering states are at the basis of the calculation of the Andreev reflection amplitudes in the presence of a non-uniform magnetization, which is described in the next section.

\section{Andreev reflection in the presence of a non-uniform magnetization direction}
\label{sec:overlap}

\subsection{Slow variations of the magnetization direction}

A spatial variation of the magnetization direction in the half metal breaks the remaining symmetries in spin space and allows for Andreev reflection at the half-metal--superconductor interface. Here we consider a continuous and slow variation of the the magnetization direction $\vm(\vr)$.
An example of such a continuous change is a domain wall, for which the net change of the magnetization angle is $\pi$, smeared out over a length $l_{\rm d}$ much larger than the microscopic lengths $k_{\rm S}^{-1}$, $k_{\uparrow}^{-1}$, and $\kappa_{\downarrow}^{-1}$. However, smaller rotation angles are possible, too, {\em e.g.} induced by strain at the interface due to lattice mismatches.\cite{kn:miao2005} 

To be specific, we choose a right-handed set of unit vectors $\ve_1$, $\ve_2$, and $\ve_3$ in spin space (which need not coincide with the coordinates used for the orbital degrees of freedom) and consider a variation of the magnetization direction $\vm$ of the form 
\begin{equation}
  \vm(\vr) = ( \ve_1 \cos \phi_{\rm m}
+  \ve_2 \sin \phi_{\rm m}) \sin \theta_{\rm m}(\vr) + 
   \ve_3\cos \theta_{\rm m}(\vr).
  \label{eq:mxyz}
\end{equation}
Such variations of the magnetization direction are sufficient to model domain walls, but they do not allow for certain continuous changes of the magnetization at a fixed polar angle, as it occurs in helical magnets. (The full expressions for arbitrary variations of $\vm$ are given in App.\ \ref{sec:appdw}.) We then employ a gauge transformation that rotates $\vm$ to the $\ve_3$-direction,
\begin{equation}
  {\cal H} \to {\cal U}(\vr)^{\dagger} {\cal H} {\cal U}(\vr),\ \
  {\cal U}(\vr) = \left( \begin{array}{cc}
  U(\vr) & 0 \\ 0 & U^*(\vr) \end{array} \right),
  \label{eq:gauge}
\end{equation}
with
\begin{equation}
\label{eq:U}
  U(\vr) = e^{i \theta_{\rm m} (\vm(\vr) \times \ve_{3}) \cdot \vsigmascript/2 \sin
    \theta_{\rm m}}.
\end{equation}
This gauge transformation adds a spin-dependent gauge potential 
\begin{eqnarray}
  \vA(\vr) &=& i \hbar U^{\dagger} \vnabla U \nonumber \\ &=&
  \frac{\hbar}{2}
  (\sigma_2 \cos\phi_{\rm m} - \sigma_1 \sin \phi_{\rm m})
  \vnabla \theta_{\rm m}  
\end{eqnarray}
to the Hamiltonian $\hat H$,\cite{kn:volovik1987} but it does not affect the singlet superconducting order parameter, since $U^{\rm T} i \sigma_2 \Delta U = i \sigma_2 \Delta$. To lowest order in the rate of change of the angle $\theta_{\rm m}$ we then find that the perturbation $\hat V$ to the Hamiltonian $\hat H$ reads
\begin{eqnarray}
  \label{eq:Vdthm}
  \hat V
  & = &
  i 
  (\sigma_2 \cos\phi_{\rm m} - \sigma_1 \sin \phi_{\rm m})
  \nonumber \\ && \mbox{} \times
  \left( \vnabla \theta \cdot \frac{\hbar^2}{2 m} \vnabla 
  + \vnabla \frac{\hbar^2}{2 m} \cdot \vnabla \theta
  \right).
\end{eqnarray}

Since we take the length scale $l_{\rm d}$ for variations of the gradient of the magnetization angle $\theta_{\rm m}$ to be large in comparison to the microscopic length scales $k_\uparrow^{-1}$, $k_{\rm S}^{-1}$ and $\kappa_{\downarrow}^{-1}$, we may neglect spatial variations of the perturbation $\hat V$ in the direction parallel to the interface. In this approximation translation symmetry along the interface is preserved (after the gauge transformation) and the scattering matrix ${\cal S}(\vkp',\vkp;\varepsilon)$ is diagonal, see Eq.\ (\ref{eq:translation}). To lowest order in the rate of change of $\theta_{\rm m}$, the Andreev reflection amplitudes may then be calculated in perturbation theory. Using the scattering states defined in the previous section, one finds
\begin{eqnarray}
  r_{\rm he}(\vkp,\varepsilon) &=& - \frac{i}{\hbar}
  \,^{\rm A}\!\langle \vkp,{\rm h },\varepsilon| {\cal V}
  | \vkp,{\rm e},\varepsilon \rangle^{\rm R}, \nonumber \\
  r_{\rm eh}(\vkp,\varepsilon) &=& - \frac{i}{\hbar}
  \,^{\rm A}\!\langle \vkp,{\rm e },\varepsilon| {\cal V}
  | \vkp,{\rm h},\varepsilon \rangle^{\rm R},
  \label{eq:perturbation}
\end{eqnarray}
where
\begin{equation}
  {\cal V} = 
  \left( \begin{array}{cc} \hat V & 0 \\ 0 & - \hat V^* \end{array} \right).
  \label{eq:Vcal}
\end{equation}

\begin{figure}
\bigskip
\epsfxsize=0.9\hsize
\epsffile{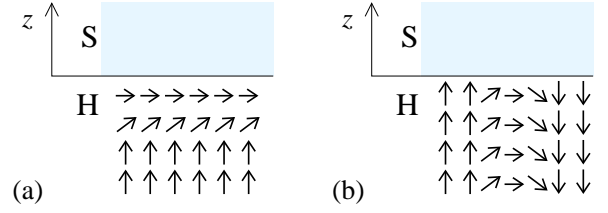}
\caption{\label{fig:3} (Color online) Illustration of a magnetization gradient perpendicular to the HS interface (a) and parallel to the HS interface (b).}
\end{figure}

We now present calculations of the Andreev reflection amplitudes for two special cases: Variation of the angle $\theta_{\rm m}$ in a direction perpendicular to the superconductor interface, as illustrated in Fig.\ \ref{fig:3}a, and variation of $\theta_{\rm m}$ in a direction parallel to the superconductor interface, which is illustrated in Fig.\ \ref{fig:3}b. These two cases differ with respect to the symmetries discussed in Sec.\ \ref{sec:symcons}: The rotation symmetry around an axis normal to the interface is preserved in the former case, whereas it is broken in the latter case. We will see that this difference has profound consequences for the Andreev reflection amplitude.

A variation perpendicular to the interface is relevant
{\em e.g.}, if the interface anisotropy at the half-metal--superconductor interface differs from the anisotropy in the bulk of the half metal. 
Calculating the Andreev reflection amplitude, we find, to leading orders in $|t|$ and $|t_{\downarrow}|$,
\begin{eqnarray}
  r_{\rm he}(\vkp;\varepsilon) &=&
  - \frac{i \varepsilon e^{-i (\phi - \phi_{\rm m})}\Delta}{\kappa_{\downarrow z}\sqrt{\Delta^2-\varepsilon^2}}
  (\vnabla \theta_{\rm m} \cdot \ve_{z})
  \nonumber \\ && \mbox{} \times
  \left[ \frac{|t t_{\downarrow}|^2}{8 \sqrt{\Delta^2-\varepsilon^2}}
  \vphantom{\left( \frac{v_{\downarrow z}^2 - v_{\uparrow z}^2}{v_{\downarrow z}^2 + v_{\uparrow z}^2} \right)}
  \right. \nonumber \\ && ~~~ \left. \mbox{} 
  + 
  \frac{(v_{\downarrow z}^2 - v_{\uparrow z}^2)
  \mbox{Re}\, t t_{\downarrow} }{\hbar k_{\uparrow z} 
  (v_{\downarrow z}^2 + v_{\uparrow z}^2)
  \sqrt{v_{\uparrow z} v_{\downarrow z}}} \right].
  ~~~~
  \label{eq:rheperpendicular}
\end{eqnarray}
The dependence of this result on the interface parameters agrees with what was derived in Sec.\ \ref{sec:symcons} using general considerations. Note that in the limit $\kappa_{\downarrow} \to \infty$, in which the minority carriers are completely expelled from the half metal, the Andreev reflection amplitude $r_{\rm he}$ vanishes. The appearance of the azimuthal magnetization angle $\phi_{\rm m}$ as a shift of the superconducting phase $\phi$ was found previously in Refs.\ \onlinecite{kn:braude2007,kn:beri2009} for the interface between a magnetic multilayer and a superconductor.

A variation of the magnetization direction in which $\vnabla \theta_{\rm m}$ is parallel to the superconductor interface breaks the rotation symmetry around an axis normal to the interface, thus allowing, in principle, for a nonzero Andreev reflection amplitude at the Fermi level $\varepsilon=0$.\cite{kn:kupferschmidt2009} Here we elaborate on our previous calculation of this effect and generalize the results of Ref.\ \onlinecite{kn:kupferschmidt2009} to the case of a finite minority wavefunction decay rate $\kappa_{\downarrow}$ in the half metal. 
Calculating the Andreev reflection amplitude according to Eq.\ (\ref{eq:perturbation}), we then find, again to leading orders in $|t|$ and $|t_{\downarrow}|$,
\begin{eqnarray}
  r_{\rm he}(\vkp;\varepsilon)
  &=&
  - \frac{e^{ - i ( \phi - \phim) }\Delta 
  }
  {\sqrt{\Delta^2 - \varepsilon^2}}(\vnabla \theta_{\rm m} \cdot \vkp)
  \nonumber \\ && \mbox{} \times
  \left[
  \frac{|t|^2 }{4 k_{{\rm S}z}^2}
  + \frac{v_{\uparrow z} \mbox{Re}\, t t_{\downarrow}}
  {(k_{\uparrow z}^2 + \kappa_{\downarrow z}^2) \sqrt{v_{\uparrow z} v_{\downarrow z}}}   
\right].
  \label{eq:rheparallel}
\end{eqnarray}
The first term in Eq.\ (\ref{eq:rheparallel}) comes from the overlap integral in Eq.\ (\ref{eq:perturbation}) inside the superconductor, whereas the second term comes from the overlap integral in the half metal. The existence of a finite contribution to $r_{\rm he}$ from inside the superconductor is responsible for the fact that $r_{\rm he}$ remains nonzero in the limit $\kappa_{\downarrow} \to \infty$ if $\theta_{\rm m}$ varies parallel to the interface. In both cases, the amplitude for the conversion of holes into electrons is given by
$$
  r_{\rm eh}(\vkp;\varepsilon) = r_{\rm he}(-\vkp;-\varepsilon)^*,
$$%
as discussed in Sec.\ \ref{sec:symcons}. For a general direction of $\vnabla \theta_{\rm m}$, the Andreev reflection amplitudes are the sums of the contributions of Eqs.\ (\ref{eq:rheperpendicular}) and (\ref{eq:rheparallel}) above.

We note that the presence of Andreev reflection at the Fermi energy for a domain wall with $\vnabla \theta_{\rm m}$ parallel to the interface is accompanied by a nontrivial angle dependence of the Andreev reflection amplitudes $r_{\rm eh}$ and $r_{\rm he}$: If $\vnabla \theta_{\rm m}$ is parallel to the interface, $r_{\rm he}$ and $r_{\rm eh}$ are even functions of $\varepsilon$, but odd functions of $\vkp$. On the other hand, if $\vnabla \theta_{\rm m}$ is normal to the interface, $r_{\rm he}$ and $r_{\rm eh}$ are odd functions of $\varepsilon$, but even functions of $\vkp$. Similar behavior has been noticed previously for the anomalous Green functions.\cite{kn:eschrig2007,kn:eschrig2008}

\subsection{ Spin-active interfaces} 

As a second example of a spatially varying magnetization direction, we now investigate a simplified model of a thin ferromagnetic or half-metallic layer located at the interface, the magnetization of which is misaligned with respect to the bulk of the half-metal. In this model of what is more generally referred to as a ``spin-active interface'',\cite{kn:eschrig2003,kn:asano2007a,kn:asano2007b,kn:eschrig2008,kn:kalenkov2009} we take the magnetization direction to be the unit vector $\ve_3$ in the entire half metal and consider a perturbation to the Hamiltonian $\hat H$ of the form
\begin{equation}
  \hat V = \tilde h \tilde \vm \cdot \hat \vsigma  \delta(z),
\end{equation}
where
\begin{equation}
  \tilde \vm  = ( \ve_1 \cos \phi_{\rm m} 
+  \ve_2 \sin \phi_{\rm m}) \sin \theta_{\rm m} + 
   \ve_3\cos \theta_{\rm m}.
\end{equation}
Spin-flip Andreev reflection at such ``spin-active'' HS interfaces has been considered previously in Refs.\ \onlinecite{kn:eschrig2008,kn:beri2009,kn:galaktionov2008}. Using Eq.\ (\ref{eq:perturbation}) to calculate the Andreev reflection amplitude to first order in $\tilde h$ and taking the limit of a tunneling interface, $|t t_{\downarrow}|^2 \ll 1$, we then find
\begin{eqnarray}
  r_{\rm he}(\vkp;\e)
  &=& - \frac{ i \e \Delta\tilde h |tt_{\downarrow}|^2 e^{ - i ( \phi - \phim )} \sin \thm}{ 2 \hbar v_{{\downarrow}z}(\Delta^2 - \e^2)}.
  \label{eq:rhespinactive}
\end{eqnarray}
The proportionality to the square of the tunneling probability is in agreement with the general considerations of Sec.\ \ref{sec:symcons}. The opposite limit of an ideal interface ($|t| = |t_{\downarrow}| = 1$) was considered in Ref.\ \onlinecite{kn:beri2009}.

\section{Thin film geometry}
\label{sec:wg}

A modification of the scattering problem arises in a lateral contact geometry as in Fig.\ \ref{fig:1}a. For this geometry a formulation with scattering states describing quasiparticles moving in the plane of the half metallic film is more relevant than the formulation of the previous sections in terms of scattering states of quasiparticles incident on the HS interface.

\begin{figure}
\bigskip
\epsfxsize=0.8\hsize
\epsffile{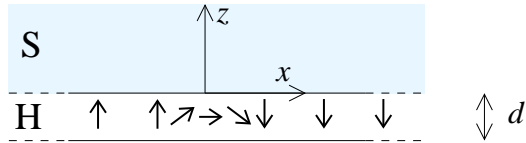}
\caption{\label{fig:4} (Color online)
Illustration of the thin-film geometry. The coordinates are chosen such that the magnetization direction does not change in the $y$ direction (perpendicular to the plane of the drawing).}
\end{figure}

The geometry we consider is shown in Fig.\ \ref{fig:4}. As before, the half-metal--superconductor interface is the plane $z=0$. The superconductor occupies the half space $z > 0$, whereas the half metallic film is in the region $-d < z < 0$. The Hamiltonian is given by Eq.\ (\ref{eq:BdG}), and hard-wall boundary conditions are applied at $z = -d$.  We choose our coordinates such, that (locally) the magnetization direction does not depend on the coordinate $y$, so that the wavevector component $k_y$ is conserved, and consider the scenario that the magnetization direction $\vm$ has a spatial non-uniformity only in a bounded region near $x = 0$. As in the previous sections, we consider the limit of a tunneling interface, and present our results to leading order in the transmission probability.

We first construct the scattering states in the presence of a uniform magnetization direction $\vm = \ve_3$. The scattering states are normalized to unit flux in the $x$ direction. 
There are electron-like and hole-like scattering states $\ket{k_yn_zs;{\rm e}}$ and $\ket{k_yn_zs;{\rm h}}$, each labeled by the conserved wavevector component $k_y$ and by integers $n_z$ and $s$, where $n_z=1,2,\ldots$ represents the quantized transverse modes in the $z$ direction and $s=\pm 1$ indicates scattering states propagating in the positive ($s=1$) and negative ($s=-1$) $x$ direction. In the limit $\kappa_{\downarrow} d \gg 1$ the hard-wall boundary conditions at $z=-d$ are inconsequential for the minority carriers, and the corresponding wavefunctions read,
\begin{eqnarray}
  \Psi_{k_yn_zs,{\rm e}}(\vr) 
  &=& 
  -i \sqrt{\frac{v_{\uparrow z}}{2 v_{\uparrow x} W_y d}}
  \Psi_{\vkp,{\rm e}}(z) e^{i k_x(\varepsilon) s x + i k_y y}, 
  \nonumber \\
  \Psi_{k_yn_zs,{\rm h}}(\vr) 
  &=& 
  i \sqrt{\frac{v_{\uparrow z}}{2 v_{\uparrow x} W_y d}}
  \Psi_{\vkp,{\rm h}}(z)
  e^{-i k_x(-\varepsilon) s x + i k_y y}, \nonumber \\
  \label{eq:psixy}
\end{eqnarray}
where $\Psi_{\vkp,{\rm e}}(z)$ and $\Psi_{\vkp,{\rm h}}(z)$ are given in Eq.\ (\ref{eq:psik}),
\begin{eqnarray}
  k_x(\varepsilon) &=& \sqrt{k_\uparrow^2-k_{\uparrow z}^2-k_y^2} + \varepsilon/\hbar v_{\uparrow x},
\end{eqnarray}
and the allowed wavenumbers $k_{\uparrow z}$ are determined from the condition that the wavefunctions vanish at $z = -d$, which implies the condition (compare with Eq.\ (\ref{eq:ccddeq}))
\begin{equation}
  2 k_{\uparrow z,{\rm e}} d + \pi - \mbox{arg}\, 
  \left( r + \frac{t^2}{e^{2 i \eta} r_{\downarrow}' - r'} \right) =
  2 \pi n_z, 
\end{equation}
for electron-like scattering states, and analogously for hole-like states
\begin{equation}
  2 k_{\uparrow z,{\rm h}} d + \pi - \mbox{arg}\, 
  \left( r^* + \frac{t^{*2}}{e^{2 i \eta} r_{\downarrow}'^* - r'^*} \right) =
  2 \pi n_z.
\end{equation}
In the limit of a tunneling interface, one has $r \to -1$ and $|t| \ll 1$, so that these conditions simplify to
\begin{equation}
  k_{\uparrow z} = \frac{\pi n_z}{d} + {\cal O}(|t|),\ \ n_z = 1,2,\ldots
\end{equation}
and the wavefunctions of the scattering states become
\begin{eqnarray}
  \Psi_{k_yn_zs,{\rm e}}(\vr) &=&
  \sqrt{\frac{2}{v_{\uparrow x} W_y d}}
  e^{i k_x(\varepsilon) s x + i k_y y}
  \sin \frac{n_z \pi z}{d}
  , \nonumber \\
  \Psi_{k_yn_zs,{\rm h}}(\vr) &=&
  \sqrt{\frac{2}{v_{\uparrow x} W_y d}}
  e^{-i k_x(-\varepsilon) s x + i k_y y}
  \sin \frac{n_z \pi z}{d}
  . \nonumber \\
\end{eqnarray}
For the purposes of our calculations higher-order terms in $|t|$ need to be kept in order to find a finite overlap between the different scattering states. In Eq.\ (\ref{eq:psixy}) we imposed periodic boundary conditions with period $W_y$ in the $y$ direction in order to ensure a proper normalization of the scattering states. The normalization factors in Eq.\ (\ref{eq:psixy}) are valid in the limit of a tunneling interface only. 

We now consider the effect of a region in which the magnetization direction is not spatially uniform. In this case, scattering between the electron-like and hole-like quasiparticle states is possible. 
To lowest order in the rate of change of the magnetization direction, the Andreev reflection amplitudes $r^{\xy}_{\rm he}$ and $r^{\xy}_{\rm eh}$ for a quasiparticle incident on a region of nonuniform magnetization can be calculated in perturbation theory as
\begin{eqnarray}
  r_{\rm he}^{\xy}(k_y;\varepsilon;n_z's';n_zs) &=& 
  - \frac{i}{\hbar}
  \langle k_yn_z's';{\rm h}| {\cal V} | k_yn_zs;{\rm e}\rangle, \nonumber \\
  r_{\rm eh}^{\xy}(k_y;\varepsilon;n_z's';n_zs) &=& 
  - \frac{i}{\hbar}
  \langle k_yn_z's';{\rm e}| {\cal V} | k_yn_zs;{\rm h}\rangle, ~~~
  \label{eq:rxydef}
\end{eqnarray}
where ${\cal V}$ is given in Eqs.\ (\ref{eq:Vdthm}) and (\ref{eq:Vcal}) above. With these equations, the problem of calculating Andreev reflection coefficients is brought into a form similar to that of the previous section. 

In principle, the perturbation ${\cal V}$ depends on the coordinate $z$, even if the magnetization angle $\theta_{\rm m}$ does not. [This can be seen, e.g., from Eq.\ (\ref{eq:Vdthm}).] Because of this, the scattering matrix is not diagonal in the transverse mode index $n_z$. However, if the length scale $l_{\rm d}$ for variations of the gradient $\vnabla \theta_{\rm m}$ in the $x$ direction (i.e., in the plane of the thin film) is large in comparison to the film thickness $d$, the resulting conservation of the momentum components $k_x$ and $k_y$ up to shifts of order $1/l_{\rm d} \ll 1/d$, together with energy conservation, constrains the possible values of the transverse momentum $k_z$. If, in addition, the film thickness $d$ is much smaller than the superconducting coherence length $\xi_{\rm S} = \hbar v_{\rm S}/\Delta$, one then finds that the off-diagonal elements of the Andreev reflection amplitude $r_{\rm he}^{\xy}(k_y;\varepsilon;n_z's';n_zs)$ are much smaller than the diagonal elements, so that one may consider $r_{\rm he}^{\xy}(k_y;\varepsilon;n_z's';n_zs)$ to be diagonal in the mode indices $n_z$ and $n_z'$. This is the case that we consider in the remainder of this section.

As a first example, we consider the case that the magnetization direction $\vm$ has the spatial dependence (\ref{eq:mxyz}) with $\theta_{\rm m}$ a function of $x$ only, so that the magnetization gradient is parallel to the HS interface. In that case one finds
\begin{eqnarray}
  r_{\rm he}^{\xy}(k_y;\varepsilon;n_z',s';n_z,s) &=&
  \delta_{n_z,n_z'} \delta_{s,-s'}
  \frac{v_{\uparrow z}}{2 v_{\uparrow x} d}
  \nonumber \\ && \mbox{} \times
  \int dx r_{\rm he}(x;\vkp;\varepsilon)
  e^{2 i \varepsilon x/\hbar v_{\uparrow x}}, \nonumber \\
  \label{eq:rhexy}
\end{eqnarray}
where $r_{\rm he}(x;\vkp;\varepsilon)$ is the Andreev reflection amplitude of Eq.\ (\ref{eq:rheparallel}), evaluated with the derivative $d \theta_{\rm m}/dx$ at position $x$. The prefactor $v_{\uparrow z}/2 v_{\uparrow x} d$ in Eq.\ (\ref{eq:rhexy}) expresses the geometric enhancement of the reflection amplitude from the coherent superposition of multiple reflections at the half-metal--superconductor interface.\cite{kn:kupferschmidt2009} The complex exponential factor accounts for the phase differences acquired by electrons and holes between these reflections. The in-plane wavevector $\vkp = s k_x(0) \ve_x + k_y \ve_y$; the small difference between the wavenumber $k_x(\varepsilon)$ of the incoming electron and the wavenumber $k_x(-\varepsilon)$ of the Andreev reflected hole is inconsequential for the calculation of $r_{\rm he}$ in the Andreev approximation $k_{\rm S} \xi_{\rm S} \gg 1$. 

If the region in which the magnetization direction varies has a size smaller than the superconducting coherence length $\xi_{\rm S}$, the $x$ dependence of the complex exponential $e^{2 i \varepsilon x/\hbar v_{\uparrow x}}$ in Eq.\ (\ref{eq:rhexy}) can be neglected. Since $r_{\rm he}(x;\vkp)$ is proportional to $d \theta_{\rm m}/dx$, the integral over $x$ gives the total change $\delta \theta_{\rm m}$ of the magnetization angle $\theta_{\rm m}$, and one finds\cite{kn:kupferschmidt2009}
\begin{eqnarray}
  \lefteqn{
  r_{\rm he}^{\xy}(k_y;\varepsilon;n_z's';n_zs)} \nonumber \\ &=&
  - \delta_{n_z,n_z'} \delta_{s,-s'}
  \frac{v_{\uparrow z} e^{-i(\phi-\phi_{\rm m})} s k_x(0) \Delta}
  {2 v_{\uparrow x} d \sqrt{\Delta^2-\varepsilon^2}}\, 
  \delta \theta_{\rm m} 
 \nonumber \\ && \mbox{} \times
  \left[ \frac{|t|^2}{4 k_{{\rm S}z}^2} +
  \frac{v_{\uparrow z} \mbox{Re}\, t t_{\downarrow}}{(k_{\uparrow z}^2 + \kappa_{\downarrow z}^2) \sqrt{v_{\uparrow z} v_{\downarrow z}}}
  \right].
  \label{eq:rhexylimit}
\end{eqnarray}
In this limit, the Andreev reflection amplitude no longer depends on the size of the domain wall, nor on the precise $x$ dependence of the magnetization angle $\theta_{\rm m}$. In the opposite limit that the domain wall size $l_{\rm d}$ is large in comparison to the superconducting coherence length, the reflection amplitude at $\varepsilon=0$ is still given by Eq.\ (\ref{eq:rhexylimit}) above, but Andreev reflection is suppressed for excitation energies $\varepsilon$ above the Thouless energy $E_{l_{\rm d}} = \hbar v_{\uparrow x}/l_{\rm d}$ of the domain wall. The precise functional form of the suppression depends on the domain wall profile. A common domain wall profile is \cite{kn:ohandley2000}
\begin{equation}
  \theta_{\rm m}(x) = \arctan[\sinh(\pi x/l_{\rm d})].
\end{equation}
In this case the suppression at finite bias $\varepsilon = e V$ is by a factor $1/\cosh(\varepsilon l_{\rm d}/\hbar v_{\uparrow x})$.

Similar to Eq.\ (\ref{eq:rhexy}) above, one finds the same 
result for variation of the magnetization along the direction normal to 
the superconducting interface, the $z$ direction, as well as for a spin active interface. In 
these cases, $r_{\rm he}(x,y;\vkp)$ in Eq.\ (\ref{eq:rhexy})  is the
Andreev reflection amplitude of Eq.\ (\ref{eq:rheperpendicular}) and Eq.\ (\ref{eq:rhespinactive}), respectively.


\section{Green functions}
\label{sec:green}

Most of the existing theoretical literature on the superconducting proximity effect in ferromagnets and half metals makes use of the Green function approach, with or without quasiclassical approximation. In the Green function approach, the induced superconducting correlations in the half metal are characterized using the ``anomalous Green function''. The symmetries of this Green function in the spin, orbital and, in particular, frequency domains are used to classify the various forms of the proximity effect.\cite{kn:bergeret2001,kn:bergeret2005,kn:eschrig2007,kn:eschrig2008} In this section, we investigate the relation between the scattering approach used here and the Green function approach. (The symmetry in the frequency domain does not play an important role for the scattering approach, because all information is encoded in ``retarded'' structures in the scattering approach.)

The fundamental equation for the Green function ${\cal G}(\vr,\vr';i \omegamatsubara)$ is the Gorkov equation
\begin{eqnarray}
  (i \omegamatsubara - {\cal H}) {\cal G}(\vr,\vr';i \omegamatsubara) =
  \delta(\vr-\vr'), \label{eq:gorkov}
\end{eqnarray}
where ${\cal H}$ is the Bogoliubov-de Gennes Hamiltonian of Eq.\ (\ref{eq:BdG}) and $\omegamatsubara$ the Matsubara frequency. Like the Hamiltonian ${\cal H}$, the Green function ${\cal G}$ has a $4 \times 4$ matrix structure, corresponding to the electron/hole and majority/minority spin degrees of freedom. The anomalous Green function is the electron-hole (eh) block of ${\cal G}$. In a half metal, only the majority component is relevant, so one has
\begin{equation}
  F(\vr,\vr';i \omegamatsubara) =
  {\cal G}_{{\rm e}\uparrow,{\rm h}\uparrow}(\vr,\vr';i \omegamatsubara).
\end{equation}

We first calculate the anomalous Green function $F(\vr,\vr';i \omegamatsubara)$ in the vicinity of the interface between a semi-infinite half metal and a superconductor. This is the geometry considered in Sec.\ \ref{sec:overlap}. In order to calculate $F(\vr,\vr';i \omegamatsubara)$ we Fourier transform the Gorkov equation (\ref{eq:gorkov}) to the coordinates $x$ and $y$, and solve the remaining one-dimensional problem using the solutions of the Bogoliubov-de Gennes equation calculated in Sec.\ \ref{sec:overlap}. For $\Omega > 0$ and coordinates $z$, $z' < 0$ ({\em i.e.,} for coordinates inside the half metal), this procedure expresses the anomalous Green function $F(\vr,\vr';i \Omega)$ in terms of the (outgoing) electron component of the exact hole-like retarded scattering state, and one finds
\begin{widetext}
\begin{eqnarray}
  F(\vr,\vr';i \Omega) &=&
  \frac{1}{(2 \pi)^2}
  \int dk_x dk_y \frac{e^{-i k_z (z-z') + i k_x (x-x') + i k_y(y-y') - |\Omega|(|z|+|z'|)}}{i \hbar v_{\uparrow z}}
  r_{\rm eh}(\vk; i \omegamatsubara),\ \
  \mbox{$\Omega > 0$},
\end{eqnarray}
where 
\begin{equation}
  k_z = \sqrt{k_{\uparrow}^2 - k_x^2 - k_y^2}.
\end{equation}
Similarly, for $\omegamatsubara < 0$, the anomalous Green function $F(\vr,\vr';i \Omega)$ is found to be proportional to the (incoming) electron component of the exact hole-like advanced scattering state,
\begin{eqnarray}
  F(\vr,\vr';i \omegamatsubara) &=&
  - \frac{1}{(2 \pi)^2 \hbar}
  \int dk_x dk_y
  \frac{e^{i k_z (z-z') + i k_x (x-x') + i k_y(y-y') - |\Omega|(|z|+|z'|)}}{i \hbar v_{\uparrow z}} r_{\rm he}(\vkp;i |\omegamatsubara|)^*,\ \
  \mbox{$\Omega < 0$}.
\end{eqnarray}
These expressions can be cast in the form
\begin{equation}
  F(\vr,\vr';i \omegamatsubara) =
  \frac{1}{2 (2 \pi)^2 i}
  \int d\vk 
  e^{i \vk \cdot (\vr-\vr')}
  f(\vk,\vR;i \omegamatsubara) \delta(\varepsilon_k - \pot_{{\rm H}\uparrow}),
  \label{eq:Ff}
\end{equation}
where $\vR = \frac{1}{2}(\vr+\vr')$, 
$\varepsilon_k = \hbar^2 k^2/2 m_{\rm H}$, and
\begin{eqnarray}
  f(\vk,\vR;i \omegamatsubara) &=& 2
  e^{-2 |\omegamatsubara| |\vR \cdot \ve_z|/\hbar v_{\uparrow z}}
  \times
  \left\{ \begin{array}{ll}
  r_{\rm eh}(\vkp;i|\omegamatsubara|) &
  \mbox{if $k_z < 0$ and $\omegamatsubara > 0$}, \\
  -r_{\rm he}(\vkp;i|\omegamatsubara|)^* &
  \mbox{if $k_z > 0$ and $\omegamatsubara < 0$}, \\
  0 & \mbox{otherwise}.
  \end{array} \right.
\end{eqnarray}
The function $f$ may be identified with the anomalous Green function in the quasiclassical theory.\cite{kn:rammer1986} Note that $f$ is nonzero only if the wavevector $\vk$ points away from the superconductor interface if $\omegamatsubara > 0$ (corresponding to a hole moving towards the superconductor, retarded case), or if $\vk$ points towards the superconductor if $\omegamatsubara < 0$ (corresponding to an electron moving towards the superconductor, advanced case).

Because of the delta function in Eq.\ (\ref{eq:Ff}), the function $f$ is meaningful for wavevectors $\vk$ with $|\vk|=k_{\uparrow}$ only.
Following the literature, we analyze moments of $f$, taken with respect to its angular dependence on $\vk$, and determine whether these are even or odd functions of the Matsubara frequency $\omegamatsubara$. The moments are defined through the relation
\begin{equation}
  f(\vk,\vR;i\omegamatsubara) =
  \sum_{l,m} Y_{lm}(\ve_{\vk}) f_{lm}(\vR;i \omegamatsubara),
\end{equation}
where $Y_{lm}$ is a spherical harmonic and $\ve_{\vk} = \vk/k$ the unit vector in the direction $\vk$. We restrict the discussion below to $l=0$ and $l=1$.

In order to determine the parity of the moments $f_{lm}$, we note that
$$
  r_{\rm eh}(\vkp,i|\omegamatsubara|) =
  r_{\rm he}(-\vkp,i|\omegamatsubara|)^*,
$$%
see Sec.\ \ref{sec:symcons}. 
If the magnetization gradient is perpendicular to the half-metal--superconductor interface, $r_{\rm he}(-\vkp,i|\omegamatsubara|)$ is an odd function of frequency, proportional to $\omegamatsubara$ for small frequencies, but an even function of $\vkp$, see Eq.\ (\ref{eq:rheperpendicular}). Hence $f_{00}$ is an odd function of $\omegamatsubara$, proportional to $\omegamatsubara$ for small frequencies.\cite{kn:eschrig2007,kn:eschrig2009} The moment $f_{10}$ is an even function of $\omegamatsubara$, proportional to $|\omegamatsubara|$ for small $\omegamatsubara$. The functions $f_{1m}$ with $m=\pm 1$ are both zero.
If the magnetization gradient is parallel to the half-metal--superconductor interface, $r_{\rm he}(-\vkp,i|\omegamatsubara|)$ is an even function of frequency, with a finite value for $\omegamatsubara \to 0$, but an odd function of $\vkp$, see Eq.\ (\ref{eq:rheparallel}). This implies that $f_{00}$ and $f_{10}$ are both zero, whereas $f_{1m}$ with $m=\pm 1$ are even functions of $\omegamatsubara$, with a finite nonzero value in the limit $\omegamatsubara \to 0$. 

The moments $f_{lm}$ calculated via the scattering approach obey the general symmetries imposed by the Pauli principle: For proximity effects induced by an $s$-wave, spin-singlet superconductor, the anomalous Green function $f_{lm}(\Omega)$ in a half metal is an odd function of $\Omega$ for even $l$ and an even function of $\Omega$ for odd $l$.\cite{kn:eschrig2007,kn:eschrig2008} 
Although $f_{00}(\Omega) \to 0$ for $\Omega \to 0$ in the two cases discussed above, the requirement that the $s$-wave amplitude $f_{00}(\Omega)$ be an odd function of $\Omega$ does not necessarily imply that $f_{00}(\Omega)$ always vanishes in this limit. Indeed, in junctions between a superconductor and a standard (not half-metallic) ferromagnet, it is known that the triplet component of $f_{00}(\Omega)$ approaches a finite value if $\Omega \to 0$. (To be precise, the triplet component of $f_{00}(\Omega) \propto \mbox{sign}\,(\Omega)$ for small $\Omega$ for a ferromagnet, in order to satisfy the antisymmetry constraint.\cite{kn:bergeret2001,kn:eschrig2007,kn:asano2007b}) For junctions involving a half metal, an example of such a singular frequency dependence is given by the case of a lateral contact to a thin half-metallic film case, which we now discuss.  

As in Sec.\ \ref{sec:wg}, we consider the case that the magnetization direction does not depend on $y$, and that the region of nonuniform magnetization is limited to the vicinity of $x=0$. We assume that the inequalities described in the paragraph following Eq.\ (\ref{eq:rxydef}) are obeyed, so that the thin-film reflection matrix $r^{\xy}_{\rm he}$ is diagonal in the transverse-mode indices $n$ and $n'$. We calculate the anomalous Green function $F(\vr,\vr';i \omegamatsubara)$ for the case that the coordinates $\vr$ and $\vr'$ are on the positive-$x$ side of the region with non-uniform magnetization. Expanding the Green function in transverse modes, one then finds
\begin{eqnarray}
  F(\vr,\vr';i \omegamatsubara) 
  &=&
  \frac{1}{2 \pi d i}
  \sum_{n} \int d\vkp \sin \frac{n \pi z}{d} \sin \frac{n \pi z'}{d}
  e^{i \vkp \cdot (\vr-\vr')} f_n(\vkp,\vR;i \omegamatsubara)
  \delta(\varepsilon_k - \pot_{{\rm H}\uparrow}),
\end{eqnarray}
where
\begin{eqnarray}
  f_{n}(\vkp,\vR;i \omegamatsubara) &=& 2
  e^{-2 |\omegamatsubara| (\vR \cdot \ve_x)/\hbar v_{\uparrow x}}
  \times
  \left\{ \begin{array}{ll}
  r_{\rm eh}^{\xy} (k_y;i|\omegamatsubara|;n+;n-) &
  \mbox{if $k_x > 0$ and $\omegamatsubara > 0$}, \\
  -r_{\rm he}^{\xy}(k_y;i|\omegamatsubara|;n+;n-)^* &
  \mbox{if $k_x < 0$ and $\omegamatsubara < 0$}, \\
  0 & \mbox{otherwise}.
  \end{array} \right.
\end{eqnarray}
\end{widetext}
The moments $f_{lm}$ with $m + l$ odd vanish in the thin film geometry because they are odd in $k_z$. Hence, we expand $f_n(\vkp,\vR;i \omegamatsubara)$ in moments as
\begin{equation}
  f_{n}(\vkp,\vR;i\omegamatsubara) =
  \sum_{l+m\ {\rm even}} Y_{lm}(\ve_k) f_{lm}(\vR;i \omegamatsubara),
\end{equation}
where $\ve_k$ is the unit vector pointing in the direction $\vkp + n \pi \ve_z/d$. In the thin-film geometry, the lowest moment $f_{00}$ is not only nonzero for magnetization gradients perpendicular to the surface, but also for magnetization gradients parallel to the surface. In both cases $f_{00}$ is an odd function of the frequency $\omegamatsubara$, but whereas $f_{00} \propto \omegamatsubara$ in the case of a magnetization gradient perpendicular to the interface, $f_{00}$ is discontinuous, $f_{00} \propto \mbox{sign}(\omegamatsubara)$, in the case that the magnetization gradient is parallel to the surface. 

The fact that the $s$-wave amplitude $f_{00}(\omega)$ may have a finite limit in the limit $\omega \to 0$ is a striking difference between the thin-film geometry and the regular geometry with a half-infinite half metal. Thus, our calculation identifies geometry as an important second factor in determining the symmetries of the anomalous Green's function --- in addition to the Pauli principle. 


\section{Conclusion}
\label{sec:conc}

In this article we have calculated the Andreev reflection amplitudes of a half-metal--superconductor (HS) junction with a spatially nonuniform magnetization direction in the half metal. General symmetry considerations enforce that the Andreev reflection amplitude $r_{\rm he}$ is zero at the Fermi level $\varepsilon=0$, except if inversion symmetry around the normal to the HS interface is broken.\cite{kn:beri2009,kn:kupferschmidt2009} On the other hand, if that is the case, $r_{\rm he}$ is an odd function of the wavevector component $\vkp$ parallel to the interface. These general results were confirmed by explicit calculations of $r_{\rm he}$ for magnetization gradients parallel and perpendicular to the interface.

Impurity scattering has not been included in the calculations presented here. As discussed in the introduction, this is not a serious shortcoming for the microscopic Andreev reflection amplitude $r_{\rm eh}$ (the amplitude for a single reflection of a quasiparticle incident on the HS interface) if the disorder is weak (mean free path $l$ much larger than the Fermi wavelengths in the superconductor or in the half metal, and than the wavefunction decay length in the half metal),
because Andreev reflection is a process that takes place on these microscopic length scales.\cite{kn:andreev1964} Hence, the microscopic reflection amplitudes $r_{\rm eh}$ of Sec.\ \ref{sec:overlap} can be used as a valid starting point for a scattering theory of a disordered HS junction. The same situation occurs at normal-metal--superconductor (NS) interfaces, where the Andreev reflection amplitudes of a clean NS interface are combined with standard theoretical methods for disordered normal metals in order to construct a theory of a disordered NS junction.\cite{kn:beenakker1995}

However, disorder has a profound effect on the type and magnitude of the induced superconducting correlations in the half metal, which are mediated by the effect of multiple Andreev reflections. This includes the calculation of the effective Andreev reflection amplitude $r_{\rm he}^{\rm eff}$ in the thin-film geometry, which represents the coherent effect of multiple Andreev reflections. In the absence of disorder, the wavevector component $\vkp$ parallel to the HS interface is conserved, and multiple Andreev reflections at the HS interface always add constructively.\cite{kn:kupferschmidt2009} With disorder, $\vkp$ is no longer conserved. Because the sign of the Andreev reflection amplitudes $r_{\rm eh}$ depends on the angle between $\vkp$ and the gradient of the magnetization angle, interference between multiple reflection events need no longer be constructive. In particular, the $p$-wave superconducting correlations induced by a magnetization gradient parallel to the HS interface are strongly sensitive to disorder, and unable to survive for large distances away from the superconductor.\cite{foot} A detailed study of disorder effects in a lateral contact between a thin half-metallic film and a superconductor is left for a future publication.

We close by noting that very recently half-metal--superconductor hybrid systems have garnered attention as possible candidates for the creation of Majorana fermion excitations,\cite{kn:sau2010} which are considered promising candidates to implement topological quantum computing.\cite{kn:dassarma2005,kn:nayak2008} While the system considered here is too elementary to be useful for quantum computation, the fundamental ingredients present here --- spin-flip scattering, half-metallicity, and $s$-wave superconducting order --- are precisely the same as those appearing in the proposals for Majorana fermion excitations.\cite{kn:sau2010} Further analysis of common and distinguishing features is thus desirable.

We gratefully acknowledge stimulating discussions with Benjamin Beri, Norman Birge, Mathias Duckheim, Francis Wilken, Dan Ralph, and Boris Spivak.
This work was supported by the Cornell Center for Materials Research under NSF Grant No.\ 0520404 and by the Alexander von Humboldt Foundation.

\begin{appendix}

\section{Explicit calculation of the gauge transformation}
\label{sec:appdw}

In this appendix we consider a general variation the magnetization direction, described by variations of both polar angles $\theta_{\rm m}$ and $\phi_{\rm m}$. If both $\theta_{\rm m}$ and $\phi_{\rm m}$ are position dependent, the transformation given in Eq.\ (\ref{eq:U}) leads to the gauge potential
\begin{eqnarray}
  \vA &=& \frac{\hbar}{2} \left[
  (\sigma_2 \cos \phi_{\rm m} - \sigma_1 \sin \phi_{\rm m})
  \vnabla \theta_{\rm m} \right. \nonumber \\ && \left. \mbox{}
  - (\sigma_1 \cos \phi_{\rm m} \sin \theta_{\rm m} 
  + \sigma_2 \sin \phi_{\rm m} \sin \theta_{\rm m})
  \vnabla \phi_{\rm m} 
  \right. \nonumber \\ && \left. \mbox{}
  - \sigma_3 (1 - \cos \theta_{\rm m}) \vnabla \phi_{\rm m}
  \right].
\end{eqnarray}
In the calculation of Sec.\ \ref{sec:overlap} only the terms proportional to $\vnabla \theta_{\rm m}$ were included. There are three terms proportional to $\vnabla \phi_{\rm m}$. Of these, the term proportional to $\sigma_3$ does not contribute to spin-flip Andreev reflection. Repeating the calculation of Sec.\ \ref{sec:overlap} with the two remaining terms proportional to $\vnabla \phi_{\rm m}$ gives the results (\ref{eq:rheperpendicular}) and (\ref{eq:rheparallel}) for gradients perpendicular and parallel to the interface, respectively, with the replacement $\vnabla \theta_{\rm m} \to \vnabla \theta_{\rm m} + i \sin \theta_{\rm m} \vnabla \phi_{\rm m}$. In the general case that the magnetization gradient has components parallel and perpendicular to the interface, the Andreev reflection amplitudes of Eqs.\ (\ref{eq:rheperpendicular}) and (\ref{eq:rheparallel}) must be added.
  
\end{appendix}


\end{document}